\documentclass[12pt,amssymb]{iopart}
\usepackage{graphicx,epsfig}
\usepackage{graphicx}
\usepackage{dcolumn}
\usepackage{bm}
\usepackage{epsfig}
\usepackage{graphics}
\usepackage{amsfonts}

\begin{document}
\title[Tunneling and transmission resonances of a Dirac particle by a double barrier] {Tunneling and transmission resonances of a Dirac particle by a double barrier}
\author{
V\'{\i}ctor M Villalba$^1$$^2$\footnote[1]{E-mail: vvillalb@scimail.uwaterloo.ca}, Luis A.
Gonz\'alez-\'Arraga$^1$}
\address{$^1$ Centro de F\'{\i}sica IVIC Apdo 21827, Caracas 1020A,
Venezuela}
\address{$^2$ Department of Physics and Astronomy, University of Waterloo, Canada}

\begin{abstract}
We calculate the tunneling process of a Dirac particle across two
square barriers separated a distance $d$,  as well as the scattering by a double  cusp
barrier where the centers of the cusps are separated a distance 
larger than their screening lengths.  Using the scattering matrix formalism, we obtain the
transmission and reflection amplitudes for the scattering  processes
of both configurations. We show that,  the presence of transmission resonances modifies the
Lorentizian shape of the energy resonances and induces the appearance of additional maxima 
in the transmission coefficient in the range of energies where transmission resonances occur. 
We calculate the Wigner time-delay and show how their maxima depend on the position of the
transmission resonance.  
\end{abstract}


\maketitle

\section{Introduction}

Barrier penetration of relativistic electrons is a very important
problem in scattering theory  and provides a
theoretical framework for different physical phenomena that are not
present in the non relativistic regime such as the Klein-paradox and
supercriticality\cite{Greiner2,Dombey,Dombey2}

The discussion of tunneling of relativistic particles by
one-dimensional potentials has been restricted to some simple
configurations such as delta potentials and square barriers, mainly
in the study of the possible relativistic corrections to mesoscopic
conduction\cite{Su} and the analysis of resonant tunneling through
multi-barrier systems\cite{Roy}. Recently\cite{Katsnelson},
electron transport through electrostatic barriers in single and
bi-layer graphene has been studied using the Dirac equation and
barrier penetration effects analogous to the Klein paradox.

The study  of transmission resonances in relativistic  wave
equations in external potentials  has been extensively discussed in
the literature\cite{Dombey,greiner,kennedy2,Jiang}. In this case we have
that, for given values of the energy and of the shape of the
barrier, the probability of transmission reaches 
unity even if the potential strength is larger than the energy of the particle, a phenomenon which is not present in the 
non relativistic case.  The relation between low momentum resonances and
supercriticality has been established by Dombey \textit{et al}\cite{Dombey,kennedy}.
Recently\cite{Jiang,kennedy,villalba1},  some results on scattering
of Dirac particles by one a dimensional potential exhibiting
resonant behavior have been reported.  

The study  of  the tunneling
effects of Dirac particles by potential barriers  has been almost
restricted to  those cases where the wave  equations are solvable in
terms of special functions and a straightforward identification of
the asymptotic states is possible. The composition of barrier 
potentials\cite{Mello}  using the
scattering matrix method permits us to consider physical
configurations  which cannot be solved in closed form. We  discuss  resonant tunneling  of
Dirac particles by a double square barrier  and  double cusp
potential  when transmission resonances are also present.  We show 
that the presence of transmission resonances close to the position of the poles of the scattering matrix 
turns out in the appearance of additional peaks in the transmission coefficient and in a modification of the Breit-Wigner 
Lorentzian profile.  We calculate
the Wigner time-delay and show that some of the peaks cannot be associated with energy resonances.

The article is structured as follows.  In Sec. 2, we calculate the 
transmission of a Dirac particle by a double barrier of equal strength in a range of energies where the 
barriers support transmission resonances. We 
calculate the Wigner time-delay of the resonances and show that,  the transmission coefficient as 
a function of energy exhibits peaks that cannot be identified as energy resonances although they 
correspond to peaks of the Wigner time-delay. In section 3, applying the composition of scattering
matrices, we discuss the transmission of a Dirac particle by a
double cusp potential. We show that when  the system exhibits energy
resonances in the range of energy where the cusps support transmission resonances the transmission coefficient and 
the Wigner-time delay  also present maxima that  cannot be identified as Breit-Wigner resonances. In Sec. 4 we 
briefly summarize our results.

\section{Double square barrier}
Resonant scattering of a Dirac particle by a square potential barrier has
been discussed by different authors in the literature \cite{Su}.
Here we are interested in studying resonant transmission of
relativistic particles by  two square barriers separated a distance
$d$   when barriers support  transmission resonances \cite{Greiner2}.

Since we are working in $(1+1)$
dimensions, we choose the following representation for the $\gamma$
matrices:
\begin{equation}
\label{rep} \gamma^{0}=i\sigma^{3}, \qquad \gamma^{1}=\sigma^{1},
\end{equation}
where $\sigma_{1}$ and $\sigma_{3}$ are the Pauli matrices. Using
the representation (\ref{rep}), the Dirac equation  in the presence
of a potential $V(x)$ takes the form\cite{villalba1}
\begin{equation}
\label{dir} [\sigma^{2}(E-eV(x))+\sigma^{1}\partial_{x}+m]\Psi=0
\end{equation}
where we have adopted the  natural units $\hbar=1$, $c=1$. 

The solution to the Dirac equation in the presence of a potential
barrier of height $V$  can be obtained after decomposing the spinor
in three regions. For $x<0$  (Region I), the solution to the Dirac
equation (\ref{dir}) has the form
\begin{eqnarray} \Psi_{I} = A_1\left(
\begin{array}{ccc}
1   \\
\frac{ik}{E-m}   \\
\end{array} \right)e^{ikx}+B_1\left(
\begin{array}{ccc}
1   \\
\frac{-ik}{E-m}   \\
\end{array} \right)e^{-ikx}\end{eqnarray}
The spinor solution in the sector II $(0<x<a)$ is
\begin{eqnarray} \Psi_{II} = \alpha\left(
\begin{array}{ccc}
1   \\
\frac{ip}{E-m-eV}   \\
\end{array} \right)e^{ipx}+\beta\left(
\begin{array}{ccc}
1   \\
\frac{-ip}{E-m-eV}   \\
\end{array} \right)e^{-ipx}\end{eqnarray}
where $p=((E-eV)^2-m^2)^{1/2}$. In the region III $x>a$ we have
\begin{eqnarray} \Psi_{II} = C_1\left(
\begin{array}{ccc}
1   \\
\frac{ik}{E-m}   \\
\end{array} \right)e^{ikx}+D_1\left(
\begin{array}{ccc}
1   \\
\frac{-ik}{E-m}   \\
\end{array} \right)e^{-ikx}\end{eqnarray}
Imposing the continuity of the spinor solution at $x=0$ and $x=a$,  we
obtain that the relation between the amplitudes of the incoming and
outgoing waves is
\begin{equation}
\label{c1}
\fl  B_1=\frac{-2i(1-\gamma^2) \sin pa}{(1-\gamma)^2
e^{ipa}-(1+\gamma)^2 e^{-ipa}}A_1-\frac{4\gamma
e^{-ika}}{(1-\gamma)^2) e^{ipa}-(1+\gamma)^2 e^{-ipa}}D_1
\end{equation}
\begin{equation}
\label{c2}
\fl  C_1=-\frac{4\gamma e^{-ika}}{(1-\gamma)^2)
e^{ipa}-(1+\gamma)^2 e^{-ipa}}A_1+\frac{-2i(1-\gamma^2) \sin pa
e^{-2ika}}{(1-\gamma)^2 e^{ipa}-(1+\gamma)^2 e^{-ipa}}D_1
\end{equation}
Using Eq. (\ref{c1}) and Eq. (\ref{c2}) we readily obtain that the
scattering matrix $S_1$ associated with a square barrier of strength
$V$ in $[0,a]$ is
\begin{eqnarray}
\fl  S_{1} = \frac{1}{(1-\gamma)^2  e^{ipa}-(1+\gamma)^2
e^{-ipa}}\left(
\begin{array}{ccc}
-2i(1-\gamma^2) \sin pa & -4\gamma e^{-ika} \\
-4\gamma e^{-ika}  & -2i(1-\gamma^2) \sin pa
e^{-2ika} \\
 \end{array}  \right) \nonumber \\
 \end{eqnarray}
where $\gamma=\frac{k(E-eV-m)}{p(E-m)}$.  Applying the property of
transformation of the scattering matrix under a translation
\cite{Mello}, we obtain that the matrix $S_{2}$ associated with a
potential barrier of strength $V_{1}$ in $[d,d+b]$ is
\begin{eqnarray}
\fl  S_{2} = \frac{1}{(1-\delta)^2  e^{iqb}-(1+\delta)^2
e^{-iqb}}\left(
\begin{array}{ccc}
-2i e^{2ikd} (1-\delta^2) \sin qb  & -4\delta e^{-ikb} \\
-4\delta e^{-ikb}  & -2i e^{-2ikd}(1-\delta^2) \sin qb e^{-2ikb} \\
 \end{array}  \right) \nonumber \\    \end{eqnarray}
where $\delta=\frac{k(E-eV_{1}-m)}{q(E-m)}$.
The composition of two scattering matrices $S_{1}$ and $S_{2}$ can
be written as\cite{Medina}
\begin{eqnarray}
\label{compose}
S=\left( \begin{array}{ccc}
r & t' \\
t & r' \\
 \end{array} \right)= \left( \begin{array}{ccc}
r_{1}+\frac{t_{1}'r_{2}t_{1}}{1-r_{1}'r_{2}} & \frac{t_{1}'t_{2}'}{1-r_{1}'r_{2}} \\
\frac{t_{2}t_{1}}{1-r_{1}'r_{2}} & r_{2}'+\frac{t_{2}r_{1}'t_{2}'}{1-r_{1}'r_{2}} \\
 \end{array} \right)
\end{eqnarray}
where the indices $1$ and $2$ correspond to the potentials $V$ and
$V_1$ respectively.  

Composing the matrices
$S_{1}$ and $S_{2}$ we obtain that the relativistic double barrier
exhibits energy resonances for the values of $E$ satisfying the
equation
\begin{eqnarray}
((1-\delta)^2 e^{iqb}-(1+\delta)^2 e^{-iqb})((1-\gamma)^2
e^{ipa}-  \nonumber \\
-(1+\gamma)^2 e^{-ipa})+4(1-\gamma^2) \sin pa (1-\delta^2)
\sin qb  e^{2ik(d-a)}=0
\end{eqnarray}

Transmission resonances for the double barrier occur when the transmission coefficient  
$T$ of the whole system is equal to unity,  this condition takes place, for $a=b$,  when $p=q$ and  $\sin(pa)=0$, 
i.e.  when  the two barriers have equal strength $V=V_{1}$ and the energy $E$ satisfies the relation:
\begin{equation}
\label{transmission}
E=V-\sqrt{n^2 \pi^2/a^2+m^2}
\end{equation}
Transmission resonances of a single barrier are maxima of the transmission coefficient, they are  are not  poles of the 
scattering matrix and therefore cannot be associated with quasibound states of the system,  nonetheless 
their presence modifies the profile and peak distribution of the transmission amplitude against the energy. 

The relativistic double barrier exhibits a discrete number of
resonances, whose position and shape reduce to those obtained in the
double delta configuration as $a\rightarrow 0$ and $b \rightarrow 0$
with $aV\rightarrow \mu$ and $bV_{1}\rightarrow \delta$ where $\mu$ and $\delta$ are the strength of the delta barriers. 
Since the 
transmission resonances for a delta barrier do not depend on the energy, all maxima of the transmission coefficient 
for the double delta are associated with energy  resonances \cite{FD89,Calogeracos}.

\begin{figure}[htbp]
\begin{center}
\vspace{1cm}
\includegraphics[width=12cm]{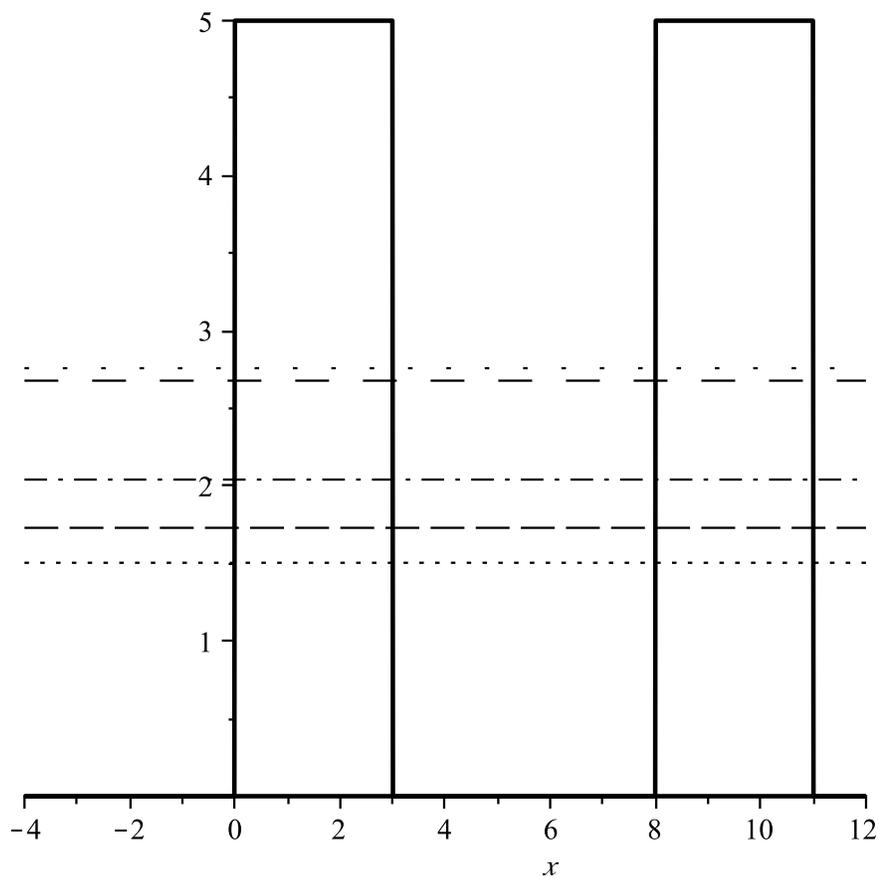}
\caption {Double barrier configuration with $V=V_1=5$, width $a=b=3$ and separated a distance $d=5$. a) The dot line
corresponds to  $\Re{E}_1=1.505$ b)   The dashed line corresponds to the transmission resonance $E_a=1.7033$ c) The dash dot 
line corresponds to $\Re{E}_2=2.0414$ d) The long dashed line corresponds to the transmission resonance $E_b=2.6791$ e) The 
space dot line corresponds to $\Re{E}_3=2.7595$. Notice that $E_b$ is close to $\Re{E}_3$} \label{ccra}
\end{center}
\end{figure}

\begin{figure}[htbp]
\begin{center}
\vspace{1cm}
\includegraphics[width=12cm]{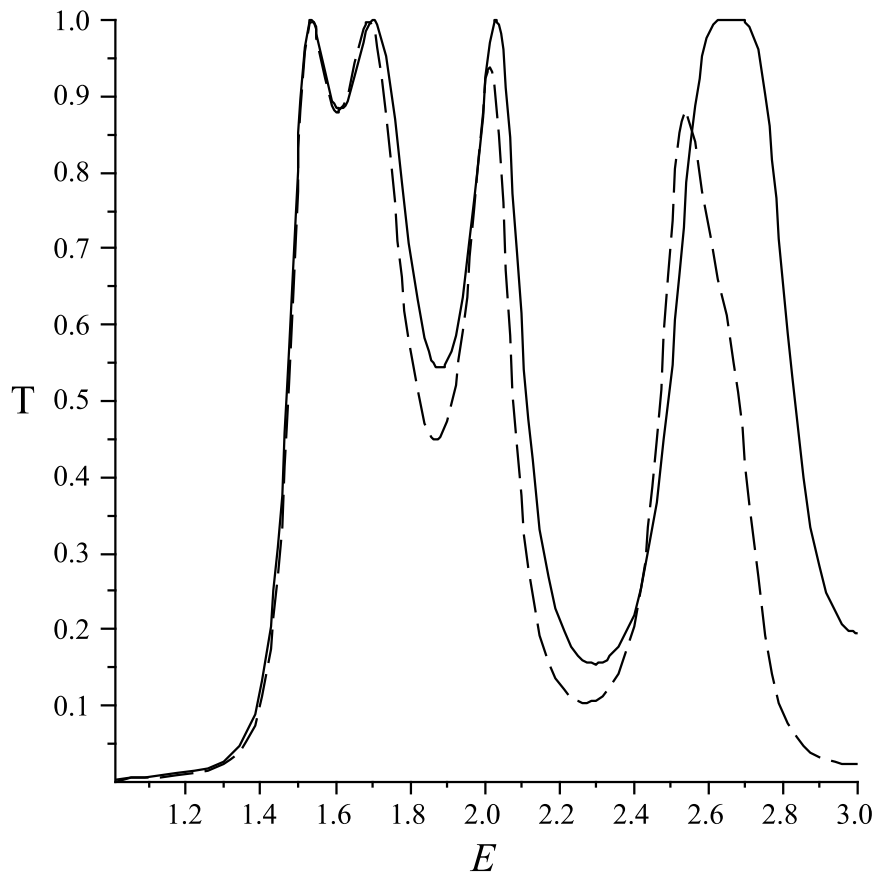}
\caption {Coefficient of transmission versus the energy
for the double barrier relativistic system. with $a=b=3$ and $d=5$. The solid line
corresponds to $V=5$ and $V_{1}=5$. 
The dashed line corresponds to $V=5$ and $V_{1}=4$} \label{ccr1}
\end{center}
\end{figure}

Fig. \ref{ccr1}  shows the behavior of the transmission coefficient
$T$ versus the energy for two square barriers with equal width $a=b=3$  and separation $d=5$. The strength of the barriers has been chosen 
to allow the presence of transmission resonances across the barriers according to Eq. (\ref{transmission}) and the mass $m$ has been equated to unity. The solid line in Fig.  \ref{ccr1} 
depicts the case with $V=V_{1}=5$  and the dashed line corresponds to $V=5$ and $V_{1}=4$. The poles of the scattering matrix $S$ for $V=V_{1}=5$
for $1<\Re(E)<3$  are $E_1= 1.5058-0.0688 I$, $E_2=2.0414-0.0810 I$, $E_3=2.7595-0.1141 I$ and the transmission resonances satisfying Eq. (\ref{transmission}) are located at $E_{a}=1.7030$ and $E_{b}=2.6791$.  Fig. \ref{ccra} depicts the double barrier configuration with $V=V_1=5$,  $d=5$ and $a=b=3$. It is also shown the position of the resonances for $\Re E<3$. From Fig.  \ref{ccr1}  it can be observed that there is a peak at the transmission 
resonance value $E_a$.  The transmission resonance at $E_b$ is located  close to $E_3$ and the peak does not exhibit a Lorentzian Breit-Wigner shape (see Fig. \ref{ccr1}). 
The dashed line corresponds to the case  $V=5$ and $V_{1}=4$. It is worth mentioning the difference we can observe between the two plots around $E_b$.

The phase shift
$\varphi$ of the transmitted amplitude $t=|t|exp(i\varphi)$ changes
rapidly near the energy resonances. The Wigner time delay
\cite{Razavy,Galindo1}
\begin{equation}
\tau=\frac{d\varphi}{dE}
\end{equation}
has maxima at positions very close to the resonance energies. The
particle is trapped for a long time in region between the barriers
before being transmitted.
\begin{figure}[htbp]
\begin{center}
\includegraphics[width=12cm]{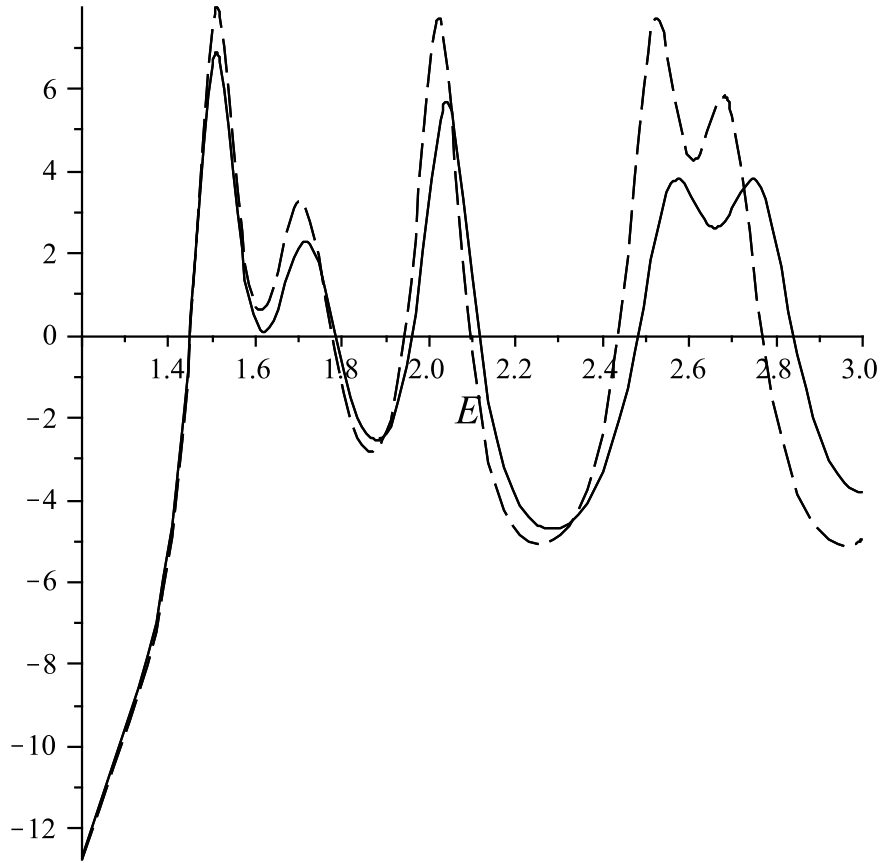}
\caption {Time delay for the relativistic double barrier with $a=b=3$ and $d=5$. The
solid line corresponds to $V=5$ and $V_{1}=5$. The dashed line corresponds to
$V=5$ and  $V_{1}=4$.}
\end{center}
\label{delayccr}
\end{figure}

Figure 3 shows the Wigner time-delay corresponding to the resonances depicted in Fig.  \ref{ccr1}   The maxima of the
Wigner time-delay are  very close to the values of $E_1$, $E_2$ and $E_3$ There are also two additional peaks at $E_a$ and 
$E_b$ that cannot be associated with  energy quasibound states but exhibiting a behavior analogous to that one observed for energy resonances. 
resonances. Fig.  \ref{delayccr} shows that the height the peaks at $E_a $ and $E_b$ is smaller than those of the energy resonances 
between $1<Re(E)<3$. 

\section{Scattering by a double cusp barrier}

In this section we proceed to study the resonant tunneling  by a double cusp potential barrier
\begin{equation}
\label{cuspp} V(x)=V_0 exp(-|x|/a)+V_1 exp(-|x-b|/d)
\end{equation}
The Dirac equation (\ref{dir}) in the presence of the potential
(\ref{cuspp}) cannot be solved exactly, therefore in order to
discuss tunneling effects we proceed to compose the scattering
matrices associated with the cusp at $x=0$ and $x=b$. This
composition gives a very good approximation to the exact result when
the separation $b$ between the centers of the cups is large in
comparison the shape parameters $a$ and $d$.

We proceed to calculate the scattering matrix $S_{1}$ associated
with the cusp potential
\begin{equation}
\label{pot1}
A^{\mu}=V_{0}exp(-\frac{\left|x\right|}{a})\delta_{0}^{\mu}
\end{equation}
We use the following representation for the Dirac matrices
\cite{villalba1}
\begin{equation}
\label{rep1} \gamma^{0}=i\sigma^{2}, \qquad \gamma^{1}=\sigma^{1},
\end{equation}
related to the matrix representation (\ref{rep}) via the unitary
transformation  $T$
\begin{equation}
T=\frac{1}{\sqrt{2}}(1-i\sigma^{1})
\end{equation}
We have that the components of the
spinor $\Psi$ solution to the Dirac equation (\ref{dir}) in the
presence of the cusp potential are
\begin{equation}\label{dirac1}
\left(
\frac{d}{dx}-i[eV_{0}exp(-\frac{\left|x\right|}{a})-E]\right)\Psi_{1}+m\Psi_{2}=0
\end{equation}
\begin{equation}\label{dirac2}
\left(
\frac{d}{dx}+i[eV_{0}exp(-\frac{\left|x\right|}{a})-E]\right)\Psi_{2}+m\Psi_{1}=0
\end{equation}
where $\Psi_{1}$ and $\Psi_{2}$ represent the upper and lower
components of $\Psi$. The solutions of the system of equations
(\ref{dirac1})-(\ref{dirac2}) can be expressed in terms of Whittaker
functions $M_{k,\mu}(z)$ \cite{abra}. We solve the Dirac equation in
the potential (\ref{pot1}) considering  an incoming plane wave as
$x\rightarrow \infty$. The potential (\ref{pot1}) induces  boundary
conditions on the solution to the Dirac equation at $x=0$. We split
the solutions to Eqs. (\ref{dirac1})-(\ref{dirac2}) into two
sectors: I $(x<0)$,  and II $(x>0)$. Looking at the asymptotic
behavior of the Whittaker function $M_{k,\mu}(z)$ as $z\rightarrow0$
\cite{abra}:
\begin{equation}
M_{k,\mu}(z)\rightarrow z^{\frac{1}{2}+\mu}exp(-\frac{z}{2})
\end{equation}
we have that the solution  behaving asymptotically as an incoming
wave from the left can be written as
\begin{equation}
\label{aa} \Psi_{inc}=\theta \left( \begin{array}{ccc}
\overline{y}^{-\frac{1}{2}} M_{k,\mu}(\overline{y}) \\
-\frac{(\frac{1}{2}+\mu+k)}{ma}\overline{y}^{-\frac{1}{2}} M_{k+1,\mu}(\overline{y})  \\
 \end{array} \right)
\end{equation}
where $\overline{y}=2iaeV_{0} exp(x/a)$. The  normalization  constant $\theta$ is
\begin{equation}
\theta=\frac{m}{(2ieaV_{0})^{\mu}\sqrt{2E(E+\sqrt{E^2-m^2})}}
\end{equation}
The spinor $\Psi_{inc}$ exhibits the following asymptotic behavior as $x\rightarrow -\infty$
\begin{equation}
\label{asinc}
 \Psi_{inc}\rightarrow\left( \begin{array}{ccc}
1 \\
-\frac{(\frac{1}{2}+\mu+k)}{ma}  \\
 \end{array} \right)(i2aeV_{0})^{\mu}exp(i\sqrt{E^2-m^2}x)
\end{equation}
The reflected wave is :
\begin{equation}
\label{ref}
\Psi_{ref}=\phi\left( \begin{array}{ccc}
\overline{y}^{-\frac{1}{2}} M_{k,-\mu}(\overline{y}) \\
-\frac{(\frac{1}{2}-\mu+k)}{ma}\overline{y}^{-\frac{1}{2}} M_{k+1,-\mu}(\overline{y})   \\
 \end{array} \right)
\end{equation}
where the normalization constant $\phi$ is
\begin{equation}
\phi=\frac{m}{(2ieaV_{0})^{-\mu}\sqrt{2E(E-\sqrt{E^2-m^2})}}
\end{equation}
the asymptotic behavior of $\Psi_{ref}$ (\ref{ref}) as
$x\rightarrow-\infty$ is:
\begin{equation}
\label{asinref}
 \Psi_{ref}\rightarrow\left( \begin{array}{ccc}
1  \\
-\frac{(\frac{1}{2}-\mu+k)}{ma}  \\
 \end{array} \right)(i2aeV_{0})^{-\mu}exp(-i\sqrt{E^2-m^2}x)
\end{equation}
The transmitted wave is :
\begin{equation}
\label{tr}
\Psi_{trans}=\left( \begin{array}{ccc}
\frac{(\frac{1}{2}-\mu+k)}{ma}y^{-\frac{1}{2}} M_{k+1,-\mu}(y) \\
y^{-\frac{1}{2}} M_{k,-\mu}(y)   \\
 \end{array} \right)
\end{equation}
where $y=2iaeV_{0}exp(-x/a)$. The asymptotic behavior of $\Psi_{trans}$ as  $x\rightarrow \infty$
takes the form:
\begin{equation}
\label{asintr}
 \Psi_{trans}\rightarrow\left( \begin{array}{ccc}
\frac{(\frac{1}{2}-\mu+k)}{ma} \\
1   \\
 \end{array} \right)(i2aeV_{0})^{\mu}exp(i\sqrt{E^2-m^2}x)
\end{equation}
which corresponds to an asymptotic plane wave traveling to the
right.

Using the  spinors $\Psi_{inc}$ (\ref{aa}), $\Psi_{ref}$ (\ref{ref}),
and $\Psi_{tr}$ (\ref{tr}) we obtain that  the solutions to the
Dirac equation in the presence of the cusp potential  in the region
I $(x<0)$ and II $(x>0)$ are

\begin{eqnarray}
\fl \Psi_{I}(x) =  a_1 \theta\left( \begin{array}{ccc}
\overline{y}^{-\frac{1}{2} M_{k,\mu}(\overline{y})} \\
-\frac{(\frac{1}{2}+\mu+k)}{ma}\overline{y}^{-\frac{1}{2} M_{k+1,\mu}(\overline{y})}   \\
 \end{array} \right)
+b_1 \phi \left( \begin{array}{ccc}
\overline{y}^{-\frac{1}{2} M_{k,-\mu}(\overline{y})} \\
-\frac{(\frac{1}{2}-\mu+k)}{ma}\overline{y}^{-\frac{1}{2} M_{k+1,-\mu}(\overline{y})}   \\
 \end{array} \right) \end{eqnarray}
\begin{eqnarray}
\fl \Psi_{II}(x) = c_1 \phi\left( \begin{array}{ccc}
\frac{(\frac{1}{2}-\mu+k)}{ma}y^{-\frac{1}{2} M_{k+1,-\mu}(y)}  \\
y^{-\frac{1}{2} M_{k,-\mu}(y)}  \\
 \end{array} \right)+ d_1 \theta\left( \begin{array}{ccc}
\frac{(\frac{1}{2}+\mu+k)}{ma}y^{-\frac{1}{2} M_{k+1,\mu}(y)} \\
y^{-\frac{1}{2} M_{k,\mu}(y)}   \\
 \end{array} \right)\end{eqnarray}
where  $a_1$, $b_1$, $c_1$, $d_1$ are constants. The continuity of the
spinor at  $x=0$ results in the following system
\begin{equation}
\fl a_1 \theta M_{k,\mu}(\nu)+b_1 \phi M_{k,-\mu}(\nu)=c_1 \phi
\frac{(\frac{1}{2}+k-\mu)}{ma}M_{k+1,-\mu}(\nu)+d_1 \theta
\frac{(\frac{1}{2}+k+\mu)}{ma} M_{k,\mu}(\nu)
\end{equation}
\begin{equation}
\fl -\frac{(\frac{1}{2}+k+\mu)}{ma}a_1 \theta
M_{k+1,\mu}(\nu)-\frac{(\frac{1}{2}+k-\mu)}{ma}b_1 \phi
M_{k+1,-\mu}(\nu)=c_1 \phi M_{k,-\mu}(\nu)+d_1 \theta M_{k,\mu}(\nu)
\end{equation}
Expressing the outgoing amplitudes  $b_1$ and  $d_1$ in terms of the
incoming amplitudes $a_1$ and $c_1$), we have:
\begin{equation}
b_1=r_{1}a_1 +t'_{1}d_1
\end{equation}
\begin{equation}
c_1=t_{1}a_1 +r'_{1}d_1
\end{equation}
where the components $r_{1}$, $r'_{1}$, $t_{1}$,  $t'_{1}$  of the matrix  $S_{1}$ for the cusp barrier are
\begin{equation}\label{menganita}
r_{1}=r'_{1}=-\frac{\theta}{\phi}\frac{\frac{(\frac{1}{2}+k)^2-\mu^{2}}{m^2a^2}M_{k+1,\mu}(\nu)M_{k+1,-\mu}(\nu)+M_{k,\mu}(\nu)M_{k,-\mu}(\nu)}{\frac{(\frac{1}{2}+k-\mu)^2}{m^2a^2}M^{2}_{k+1,-\mu}+M^{2}_{k,-\mu}(\nu)}
\end{equation}
\begin{equation}\label{sultanita}
t_{1}=t'_{1}=\frac{\theta}{\phi}\frac{\frac{(\frac{1}{2}+k+\mu)}{ma}M_{k,-\mu}(\nu)M_{k+1,\mu}(\nu)-\frac{(\frac{1}{2}+k-\mu)}{ma}M_{k,\mu}(\nu)M_{k+1,-\mu}(\nu)}{\frac{(\frac{1}{2}+k-\mu)^2}{m^2a^2}M^{2}_{k+1,-\mu}+M^{2}_{k,-\mu}(\nu)}
\end{equation}
Applying the translation property of the scattering matrix, we
obtain that the scattering matrix $S_2$ associated with the cusp
barrier $V_1 exp(-|x-b|/d)$ has the following components:

\begin{equation}
\label{menganita2}
r_{2}=-e^{i2pb}\frac{\theta^*}{\phi^*}\frac{\frac{(\frac{1}{2}+\kappa)^2-\eta^{2}}{m^2d^2}M_{\kappa+1,\eta}(\epsilon)M_{\kappa+1,-\eta}(\epsilon)+M_{\kappa,\eta}(\epsilon)M_{\kappa,-\eta}(\epsilon)}{\frac{(\frac{1}{2}+\kappa-\eta)^2}{m^2d^2}M^{2}_{\kappa+1,-\eta}+M^{2}_{\kappa,-\eta}(\epsilon)}
\end{equation}
\begin{equation}
\label{sultanita2}
r'_{2}=-e^{-i2pb}\frac{\theta^*}{\phi^*}\frac{\frac{(\frac{1}{2}+\kappa)^2-\eta^{2}}{m^2d^2}M_{\kappa+1,\eta}(\epsilon)M_{\kappa+1,-\eta}(\epsilon)+M_{\kappa,\eta}(\epsilon)M_{\kappa,-\eta}(\epsilon)}{\frac{(\frac{1}{2}+\kappa-\eta)^2}{m^2d^2}M^{2}_{\kappa+1,-\eta}+M^{2}_{\kappa,-\eta}(\epsilon)}
\end{equation}
\begin{equation}
t_{2}=t'_{2}=\frac{\theta^*}{\phi^*}\frac{\frac{(\frac{1}{2}+\kappa+\eta)}{md}M_{\kappa,-\eta}(\epsilon)M_{\kappa+1,\eta}(\epsilon)-\frac{(\frac{1}{2}+\kappa-\eta)}{md}M_{\kappa,\eta}(\epsilon)M_{\kappa+1,-\eta}(\epsilon)}{\frac{(\frac{1}{2}+\kappa-\eta)^2}{m^2d^2}M^{2}_{\kappa+1,-\eta}+M^{2}_{\kappa,-\eta}(\epsilon)}
\end{equation}
where
\begin{equation}
\kappa=iEd-\frac{1}{2}, \qquad  \eta=id\sqrt{E^2-m^2}, \qquad
\epsilon=i2dV_{1},
\end{equation}
and
\begin{equation}
\fl \theta^*=\frac{m}{(2iecV_{1})^{\eta}\sqrt{2E(E+\sqrt{E^2-m^2})}},
\qquad
\phi^*=\frac{m}{(2iedV_{1})^{-\eta}\sqrt{2E(E-\sqrt{E^2-m^2})}}.
\end{equation}
Using the expressions (\ref{menganita}), (\ref{sultanita}),
(\ref{menganita2}), and (\ref{sultanita2}) we calculate the
scattering matrix (\ref{compose}) for  the double cusp (\ref{cuspp}). Using the components
$r$ and $t$ of the $S$ matrix we calculate the energy resonances.

\begin{figure}[htbp]
\begin{center}
\vspace{1cm}
\includegraphics[width=12cm]{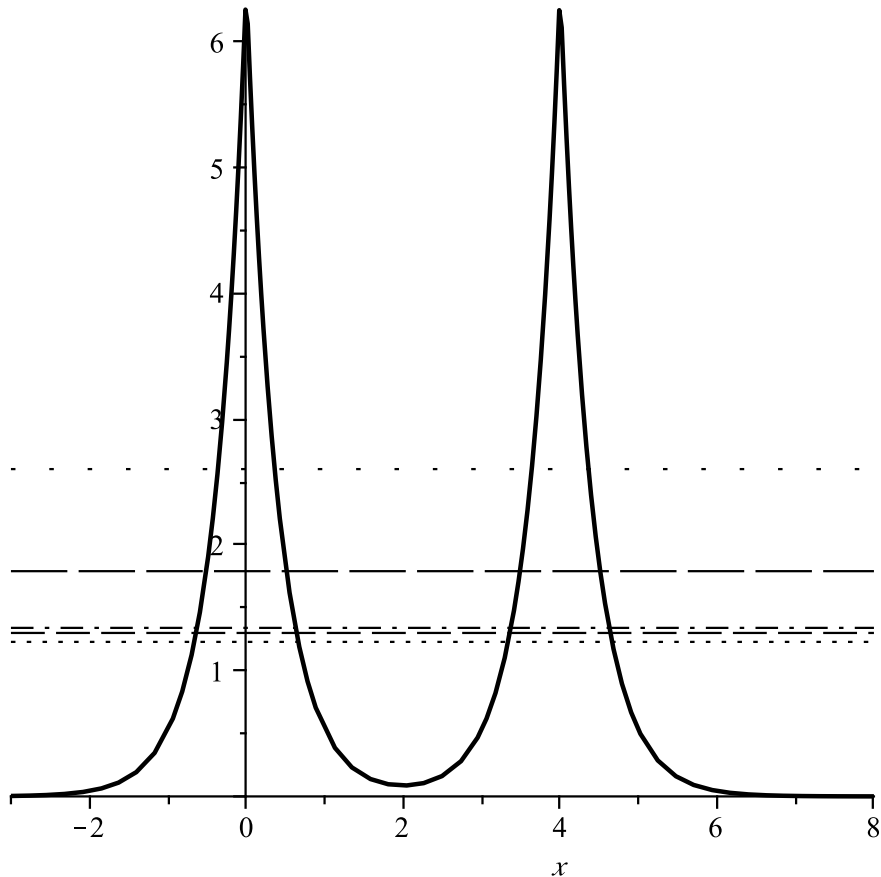}
\caption {Two cusp potentials with $a=d=0.4$,   $V=V_1=6.4271$ and with peaks separated a distance $b=4$. a) The dot line 
corresponds to $\Re{E}_1=1.229$,  b) the dashed line corresponds to the transmission resonance $E_{res}=1.3$,  c) The dash-dot 
line corresponds to $\Re{E}_2=1.340$ d) The long dashed line corresponds to $\Re{E}_3=1.791$ e) The space dot line corresponds
to $\Re{E}_4=1.603$ Notice that $E_{res}$, $\Re{E}_1$ and $\Re{E}_2$ are very close each other.} \label{ccrb}
\end{center}
\end{figure}

\begin{figure}[htbp]
\vspace{1cm}
\begin{center}
\includegraphics[width=12cm]{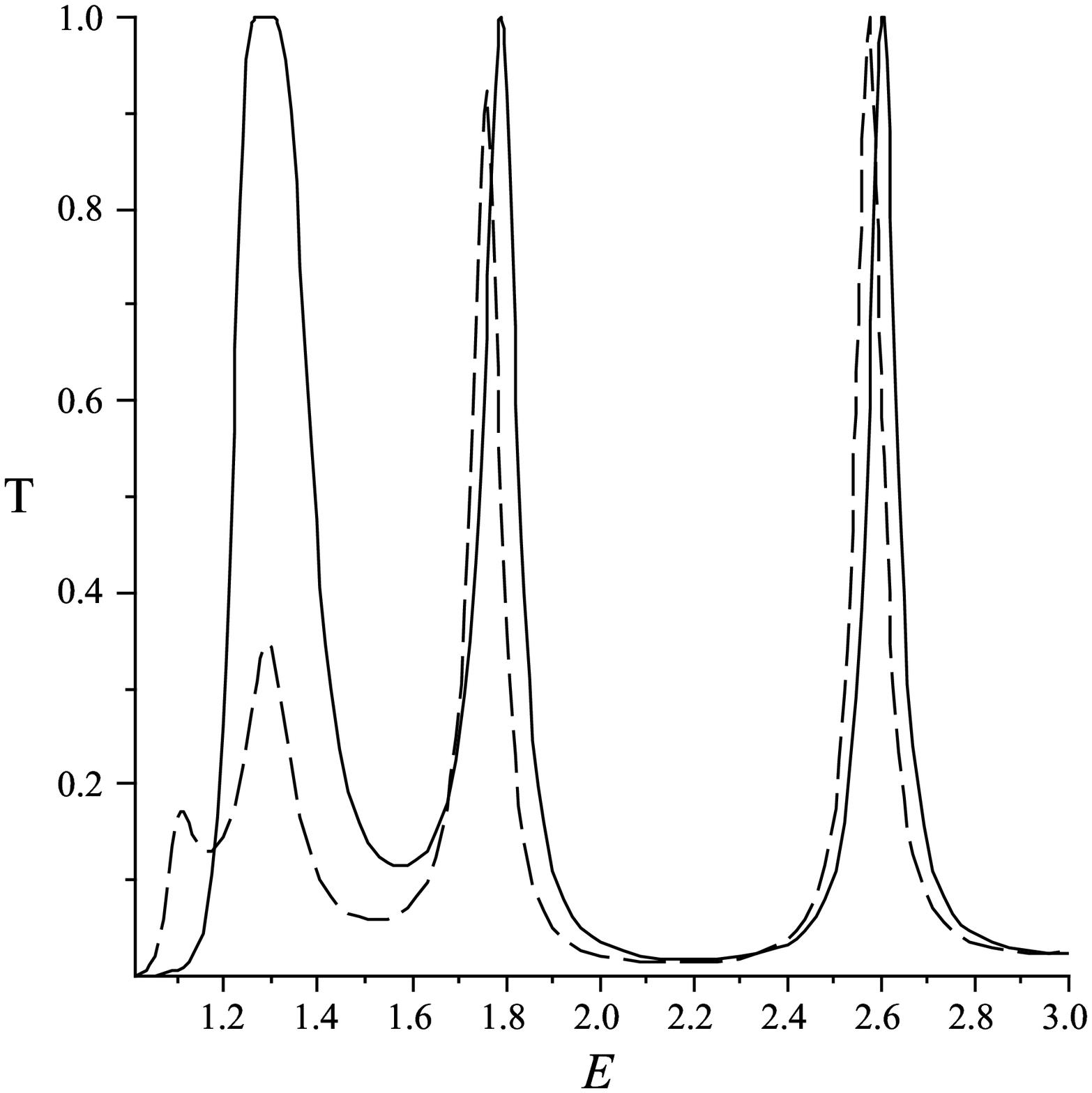}
\caption {Coefficient of transmission T versus the energy  for the
double cusp relativistic system with $a=d=0.4$ and $b=4$. The solid line corresponds to
$V_{0}=V_{1}=6.4271$.  The dashed line
corresponds  to $V_{0}=6.4271$ and $V_{1}=6$}  \label{double}
\end{center}
\end{figure}

The solid line in Fig. \ref{double} shows the behavior of the transmission coefficient 
$T$ versus the energy for the potential (\ref{cuspp}) with $a=d=0.4$ and $b=4$.  The solid line depicts de case when both cups have the same height 
$V_{0}=V_{1}=6.4271$ which is a transmission resonance value for an energy of $E_{res}=1.3$  The solid 
line has four peaks corresponding to the scattering matrix poles $E_1=1.2290-0.0451 I$, $E_2=1.3404-0.07391 I$, 
$E_3=1.79129-0.0423 I$, and $E_4=2.6031-0.0353 I$.  Fig. \ref{ccrb} depicts the double cusp configuration with $V_0=V_1=6.4271$,  $a=d=0.4$ and $b=4$. It is also shown the position of the resonances for $\Re E<3$. It should be noticed that in Fig.  \ref{double} the  peak around 1.3 does not correspond to
a single resonant value but to a a superposition of $E_1$, $E_2$ and the transmission resonance $E_{res}$ (see Fig. \ref{ccrb}) that modifies the standard Breit-Wigner profile 
of a energy resonance  \cite{Galindo1}.  The dashed line depicts the behavior the transmission coefficient against against $E$ for $V_{0}=6.4271$ and 
$V_{1}=6$. In this case there is not overlapping of  energy resonances around $E=1.3$ since there are not transmission resonances for 
cusp potentials with unequal strengths. 

Fig. 6 shows the Wigner time-delay for the  two-cusp barriers described by Fig. \ref{double}.  The solid line shows that the transmission resonance located at $E_{res}=1.3$ modifies  the shape of the peak corresponding to the energy resonance  $E_2$. The dashed line shows 
the Wigner time-delay for $V_{0}=6.4271$ and  $V_{1}=6$.

\begin{figure}[htbp]
\begin{center}
\label{delaybbr}
\includegraphics[width=12cm]{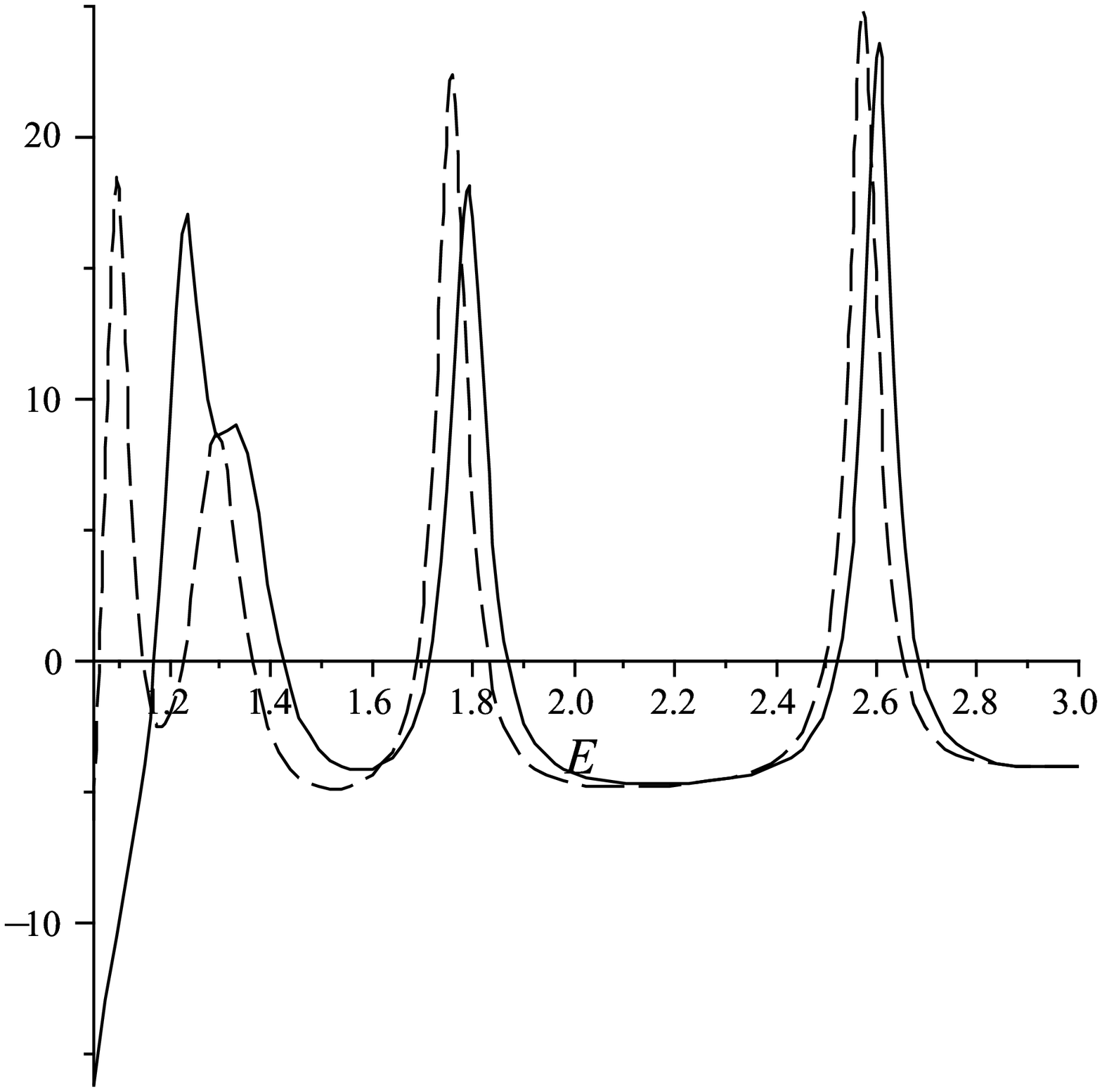}
\caption {Time delay for the relativistic double-cusp system  with $a=d=0.4$ and $b=4$. 
The solid line corresponds to
$V_{0}=V_{1}=6.4271$. The dashed line
corresponds to $V_{0}=6.4271$ and $V_{1}=6.$} 
\end{center}
\end{figure}

\section{Concluding remarks}

In the present article we have studied resonant tunneling of Dirac
particles by a  double square barrier and a double
cusp potential. We have shown that the presence of transmission resonances
associated with the square barrier or the cusp potential modifies the shape and
distribution of the energy resonances. We have shown that the Wigner time-delay of the
energy resonances is modified when they are located close to a transmission resonance. 
We have also shown that  there are no transmission resonances when the cusps or
barriers have different strengths.  The transmission with no reflection across
a barrier whose  amplitude is stronger than the energy of the tunneling  particle is a relativistic 
phenomenon with no analogue in the Schr\"odinger framework. 

The application of semianalitical methods such as the phase-integral
approximation\cite{Froman} or the uniform approximation\cite{Berry}
could be applied in  the study of transmission
resonances of composite systems where  no analytic solutions for the
potentials are available.

\end{document}